\documentclass[nofootinbib, prd]{revtex4}
\usepackage{amsmath}
\usepackage{amsfonts}
\usepackage{amssymb}
\usepackage{dsfont}
\usepackage{mathrsfs}
\usepackage{bm,slashed}
\usepackage{graphicx}
\usepackage{color}
\usepackage[normalem]{ulem}
\usepackage{hyperref}
\usepackage{MnSymbol}
\hypersetup{
    colorlinks=true, 
    linktoc=page,    
    linkcolor=blue,  
    urlcolor=black
}
\usepackage{epstopdf}
\newcommand{\ie}{{\emph{i.e.}}~}
\usepackage{color}

\newcommand{\be}{\nopagebreak[3]\begin{equation}}
\newcommand{\ee}{\end{equation}}
\newcommand{\bee}{\nopagebreak[3]\begin{equation*}}
\newcommand{\eee}{\end{equation*}}
\newcommand{\ba}{\nopagebreak[3]\begin{eqnarray}}
\newcommand{\ea}{\end{eqnarray}}
\newcommand{\baa}{\nopagebreak[3]\begin{eqnarray*}}
\newcommand{\eaa}{\end{eqnarray*}}
\newcommand{\la}{\label}
\newcommand{\n}{\nonumber}
\DeclareFontFamily{U}{rsfs}{}         
\DeclareFontShape{U}{rsfs}{m}{n}{<5> rsfs5 <6><7> rsfs7          %
  <8><9><10><10.95><12><14.4><17.28><20.74><24.88> rsfs10}{}     %
\DeclareMathAlphabet{\mathfs}{U}{rsfs}{m}{n}                     %
\newcommand{\mfs}[1]{\mathfs {#1}}                               %
\newcommand{\sL}{{\mfs L}}
\newcommand{\va}{\scriptscriptstyle}

\newcommand{\HH}{{\mathcal{H}}}
\newcommand{\lev}{\nabla}
\newcommand{\af}{{\overline{\nabla}}}
\newcommand{\lie}{{\pounds}}
\newcommand{\lc}{\gamma}

\newtheorem{definition}{Definition}

\begin{document}
\title{Spacetime thermodynamics in the presence of torsion}

\author{Ramit Dey}
\email[]{rdey@sissa.it}

\author{Stefano Liberati}
\email[]{liberati@sissa.it}

\author{Daniele Pranzetti}
\email[]{dpranzetti@sissa.it}

\affiliation{SISSA, 
Via Bonomea 265, 34136 Trieste, Italy and INFN, Sezione di Trieste}
\begin{abstract}
It was shown by Jacobson in 1995 that the Einstein equation can be derived as a local constitutive equation for an equilibrium spacetime thermodynamics. With the aim to understand if such thermodynamical description is an intrinsic property of gravitation, many attempts have been done so far to generalise this treatment to a broader class of gravitational theories. Here we consider the case of the Einstein--Cartan theory as a prototype of theories with non-propagating torsion. In doing so, we study the properties of Killing horizons in the presence of torsion, establish the notion of local causal horizon in Riemann--Cartan spacetimes, and derive the generalised Raychaudhuri equation for this kind of geometries.  Then, starting with the entropy that can be associated to these local causal horizons, we derive the Einstein--Cartan equation by implementing the Clausius equation. We outline two ways of proceeding with the derivation depending on whether we take torsion as a geometric field or as a matter field. In both cases we need to add internal entropy production terms to the Clausius equation as the shear and twist cannot be taken to be zero a priori for our setup. This fact implies the necessity of a non-equilibrium thermodynamics treatment for the local causal horizon. Furthermore, it implies that a non-zero twist at the horizon will in general contribute to the Hartle--Hawking tidal heating for black holes with possible implications for future observations.  
\end{abstract}

\maketitle
\section{\label{sec:level1}Introduction}
The analogy between the laws of black hole mechanics and the known thermodynamic laws were first shown using classical General Relativity \cite{Bekenstein:1973ur, Bardeen:1973gs}. Further studying quantum fields in a background spacetime containing a black hole, Hawking \cite{Hawking:1974sw} showed that black holes exhibit spontaneous emission and have a physical temperature, thus behaving as a true thermodynamic system. The study of black hole thermodynamics and the notion of entropy associated with the black hole horizon represents some of the best tools for gaining insights about a quantum theory of gravity.

One line of thought that shed further light into the thermodynamical nature of gravity was developed by Jacobson in \cite{Jacobson:1995ab} (see also \cite{Padmanabhan:2009vy, Padmanabhan:2014jta} for reviews of other developments on this point of view). Assuming that the entropy of a Local Causal Horizon (LCH) is proportional to its area, the Einstein equation was derived  implementing the Clausius equation. The heat flux flowing across the LCH was related to the entropy using the Clausius equation, with the temperature of the horizon  assumed to be the Unruh temperature \cite{Unruh:1976db}. This shows that the thermodynamics associated to the underlying degrees of freedom gives rise to spacetime dynamics and thus the Einstein equation can be interpreted as an equation of state. The far reaching consequence of such a result is that it might be hinting that gravity is emergent or in other words dynamics of spacetime is a manifestation of some fundamental dynamical degrees of freedom underlying the gravitational ones. The identification of these microscopic degrees of freedom as the quanta of the gravitational field is a perspective for instance pointed out in \cite{Smolin:2012ys, Chirco:2014saa} (the recovery of General Relativity from the continuum, coarse grained limit of the loop quantization of spacetime microscopic degrees of freedom using statistic mechanical methods is currently a very active line of research \cite{Gielen:2013kla, Oriti:2015qva, Oriti:2015rwa, Oriti:2016qtz, Dittrich:2013xwa, Dittrich:2014ala, Dittrich:2014mxa, Dittrich:2016tys}).

{If gravity is emergent and the dynamics is just a result of coarse graining,  then it is very important to show that this thermodynamic interpretation of gravity holds for theories beyond General Relativity. This is specially important if gravity is considered as an effective field theory \cite{Donoghue:1995cz}, where the Einstein--Hilbert action is just the first term in the action.  }
So Jacobson's result naturally led to the question whether equations of motion for more generalised theories of gravity, such as higher derivative theories, can be derived from local thermodynamical variables for a LCH by implementing the Clausius equation in the same way as it was done for General Relativity. It was possible to obtain the equation of motion for $f(R)$ theory \cite{Eling:2006aw,Chirco:2010sw} after modification of the Clausius equation to add internal entropy production terms but there were many technical hindrances for a further generalization using just the proportionality of entropy with horizon area. In  \cite{Guedens:2011dy, Dey:2016zka} a further generalization of this result for higher derivative theories of gravity was achieved by assuming the entropy to have a form similar to the Noether charge associated to the diffeomorphisms of the theory.

On the other hand, one can try to extend General Relativity by including the intrinsic spin of the particle in the geometric description of spacetime itself \cite{Hehl:1976kj}. For this one needs to introduce torsion as an additional degree of freedom for spacetime, besides the metric, and use Riemann--Cartan geometry instead of Riemannian geometry. The most well known example of such a theory is the Einstein--Cartan (EC) theory \cite{cartan1923varietes} in which the Einstein equations are replaced by the Einstein-Cartan-Sciama-Kibble field equations \cite{sciama1962analogy,kibble1961lorentz} after inclusion of torsion (for more recent reviews on spacetime torsion see, for instance, \cite{Shapiro:2001rz, Hammond:2002rm, Obukhov:2006gea, Poplawski:2009fb}). In this paper we address the question if field equations of such a theory, which includes torsion as well, can emerge out of the local thermodynamic variables of a LCH in the framework of Riemann--Cartan geometry.

 Before presenting the main results of the paper, we review in Section \ref{sec:Review} previous thermodynamical derivations of gravity equations of motion \`a la Jacobson.
 We  then focus on the geometric setup of the problem  by introducing, in Section \ref{sec:RC}, the Riemann--Cartan spacetime and  by describing, in Section \ref{sec:Inertial},  how one can have a local inertial frame along with LCH in such a framework.
 In Section \ref{sec:Hor} we analyze the horizon properties in presence of torsion and find some restriction on the torsion components in order to have a well defined notion of surface gravity.
  In Section \ref{sec:Ray} we derive the null Raychaudhuri equation in the presence of torsion. 
  In Section \ref{sec:Ent} we show that the horizon entropy change is still given by the (integral of the) expansion of the null congruence in Riemann--Cartan spacetime.
 Finally, in Section \ref{sec:TEC} we introduce the Einstein--Cartan equation and present its non-equilibrium thermodynamical derivation. As an immediate consequence, we obtain a modified shear viscosity contribution with terms proportional to the torsion tensor; possible experimental implications of this result in the context of gravitational waves emission are briefly addressed in the final Section \ref{sec:Disc}.  Further details and proofs are presented in the Appendices \ref{App:A}, \ref{App:B}, \ref{App:C}.


\section{Review of previous derivations}\la{sec:Review}
To answer the question if gravity is emergent and the known field equations for any theory of gravity are just an equation of state (i.e they are  manifestation of some local microscopic degrees of freedom), Jacobson gave a series of arguments and derived the Einstein equation starting with an entropy of a local Rindler horizon and implementing the equilibrium Clausius equation \cite{Jacobson:1995ab}. 
In this original derivation, the horizon shear was assumed to vanish. However, in an attempt to extend such thermodynamical derivation to $F(R)$ gravity,
it was later realized in \cite{Eling:2006aw, Eling:2008af, Chirco:2009dc} that this unjustified assumption can be relaxed and the LCH shear contribution to the  Raychaudhuri equation can be taken into account by considering a non-equilibrium thermodynamical setting. More precisely, the horizon shear represents an internal entropy production term, identified as a tidal heating term encoding spacetime viscosity.

We briefly review these derivations here stating the key assumptions and the necessary requirements.

 \subsection{Einsteins equation of state} \la{sec:RevE}
Using the Einsteins equivalence principle one can view the local neighborhood of any arbitrary spacetime point $p$ as flat spacetime. Through $p$ one can consider a spacelike 2-surface element $\Sigma$. The past horizon of $\Sigma$ is called the local causal horizon, which can be thought of as a localized Rindler horizon passing through the point $p$. A priori, we do not fix the values of the  affine expansion, $\theta$, and shear, $\sigma_{\mu\nu}$, of the past directed null congruence. This implies that one cannot assume
an equilibrium thermodynamics of the horizon defined within this local patch around $p$, as a non-zero expansion or shear would imply the surface (cross section of the LCH) is shearing or expanding or contracting. 

Therefore,  the Clausius equation within this local patch has to be generalized by taking into account also a internal entropy production term, i.e 
\be\la{NECla}
\delta Q=T (d S+ dS_i)\,.
\ee
One can interpret the heat as energy flow across the horizon and the equilibrium entropy term, $dS$, as the entanglement entropy of some field degrees of freedom across the horizon, which indeed results to be proportional to the area of the horizon. The second entropy term, $dS_i$, represents instead an out of equilibrium contribution.
For a uniformly accelerated observer just outside the horizon, the temperature appearing in \eqref{NECla} can be understood as the Unruh temperature, $T=\hbar/2\pi$, associated to the approximate local Rindler horizon due to boost invariance (Bisognano--Wichmann theorem \cite{Haag:1992hx}).

 For consistency, one should use the same accelerated observer to define the energy flux (or in other words the heat flow across the LCH).  
Considering a local Rindler horizon through $p$, one can assume an approximate local (boost) Killing vector, $\xi^{\mu}$, generating the horizon. To the past of $\Sigma$ the heat flux can be defined as the boost energy across the horizon
\ba
\delta Q=\int_{\HH}T_{\mu \nu} \xi^{\mu}d\Sigma^{\nu}\,,
\ea
where the integral is over the generators of the ``inside'' horizon $\HH$ of $\Sigma$. We can assume a vector $k^{\mu}$ to be tangent to the horizon and parametrized by an affine parameter $\lambda$, then we have the relation $\xi^{\mu}=-\lambda k^{\mu}$, assuming $\lambda$ is negative on the past of  $\Sigma$. We also have the relation $d\Sigma^{\mu}=k^{\mu} d \lambda dA$, where $dA$ is the area element of the cross section of the LCH. Using these relations, the heat flux is given as 
\ba \la{deltaQ}
\delta Q=- \int_{\HH}\lambda\,T_{\mu \nu} k^{\mu} k^{\nu}d\lambda dA\,.
\ea

Assuming the proportionality of entropy with area of the LCH, one can write $d S=\alpha \delta A$, with the variation of the area, $\delta A$, of the LCH cross section given by 
\ba
\delta A=\int_{\HH} \theta d\lambda dA\,,
\ea
where $\theta$ is the expansion of the null congruence. 
The UV cut-off $\alpha$ is given by the Planck length $\ell_p$, modulo a numerical constant.
We can then expand the variation of the the area
about the point $p$, corresponding to the surface at $\lambda=0$, as
\be
\theta\approx \theta_p+\lambda\left. \frac{d\theta}{d\lambda}\right|_p+\mathcal{O}(\lambda^2)
\ee
and plug it in the area variation. By means of the Raychaudhuri equation, we thus obtain
\ba \la{deltaA}
\delta A=\int_{\HH} \left.\left(\theta+\lambda\dfrac{d \theta}{d \lambda} \right)\right|_{p} d \lambda d A=\int_{\HH}  \left.\left[\theta-\lambda
\left(\frac{1}{2}\theta^2+\sigma_{\mu\nu}\sigma^{\mu\nu}
+ R_{\mu \nu} k^{\mu}k^{\nu}\right)\right]\right|_{p} d \lambda dA\,,
\ea
where  the twist is  set to zero since $k^{\mu}$ is hypersurface orthogonal. By plugging   \eqref{deltaA} and the Unruh temperature in the non-equilibrium Clausius equation \eqref{NECla}, at zeroth order in $\lambda$ one gets the condition $\theta_p=0$; then, at first order in $\lambda$, for all null $k^{\mu}$, we obtain the equations
\be\la{Clagr}
\frac{2\pi}{\hbar \alpha}T_{\mu \nu}=R_{\mu \nu}+F(x)g_{\mu\nu}
\ee
and
\be\la{dSi}
dS_i=-\alpha\int_{\HH}\left.\lambda\, \sigma_{\mu\nu}\sigma^{\mu\nu}\right|_{p}d \lambda dA\,.
\ee
Using the local conservation of the stress-energy tensor together with the Bianchi identity in \eqref{Clagr}, one obtains $F=-1/2 R + \Lambda$, thus recovering the full Einstein equation, once we set $\alpha=1/4\hbar G$.
Eq. \eqref{dSi}, on the other hand, shows how the internal entropy production contribution can be interpreted as a shear viscosity term of the stretched horizon \cite{Thorne:1986iy}.

\section{Riemann--Cartan spacetime}\la{sec:RC}

In this work, our objective is to derive the Einstein--Cartan equation in a similar thermodynamical approach as illustrated above. Since we are interested in a spacetime having torsion in the background, we give some key features of a Riemann--Cartan spacetime in this Section and more details are reviewed in Appendix \ref{App:A}. A general affine connection is parametrized by its connection coefficients.  Assuming metric compatibility, the non--Riemannian part of the connection is uniquely determined by the torsion tensor. In a generic coordinate basis, the connection coefficients read
\begin{equation}\la{ecconn}
\Gamma^{\mu}\!_{\nu\rho} = \lc^{\mu}\!_{\nu\rho} + K^{\mu}\!_{\nu\rho}\,,
\end{equation}
where, without loss of generality, we have separated the contribution of a Levi--Civita part $\gamma$ and a contortion tensor $K$ which contains the torsion properties.
We will denote the Levi--Civita covariant derivative with $\lev$, while the general one will be barred $\af$.

\begin{definition} Torsion tensor:
\begin{equation}\la{T}
T^{\mu}\!_{\nu\rho} := \Gamma^{\mu}\!_{\nu\rho} -  \Gamma^{\mu}\!_{\rho\nu}\,.
\end{equation}
\end{definition}
The contortion tensor $K$ can be rewritten in terms of the torsion tensor as
\begin{definition} Cotortion tensor:
\begin{equation}\la{K}
K^{\mu}\!_{\nu\rho} := \frac{1}{2}\left(T^{\mu}\!_{\nu\rho}-  T_{\nu\rho}\!^\mu 
-T_{\rho\nu}\!^\mu\right)\,.
\end{equation}
\end{definition}
  This definition follows from the fact that we can always split the tensor $K$ into the symmetric and antisymmetric parts
\begin{equation}
K^{\mu}\!_{\nu\rho} = U^{\mu}\!_{\nu\rho} + \frac{1}{2} T^{\mu}\!_{\nu\rho},
\label{decomp0}
\end{equation}
where $U$ is another tensor, symmetric in the two lower indices. From the metricity condition
\be
\af_{\mu}g_{\nu\rho}=0\,,
\ee
it is immediate to get
\begin{equation}\la{U}
U^{\mu}\!_{\nu\rho} =- \frac{1}{2}\left(
   T_{\nu\rho}\!^\mu 
+ T_{\rho\nu}\!^\mu
 \right)\,,
\end{equation}
where spacetime indices are raised, lowered and contracted with the metric $g_{\mu\nu}$  (for instance, $ T_{\nu\rho}\!^\mu =g_{\nu\alpha}g^{\mu\beta}T^\alpha\!_{\rho\beta}$).

Let us introduce the modified torsion tensor.

\begin{definition} Modified torsion tensor:
\be\la{S}
S^\mu\!_{\nu\sigma}:= T^\mu_{\nu\sigma}+ T_\sigma \delta^\mu_\nu-T_\nu \delta^\mu_\sigma \,,
\ee
\end{definition}
where
\be
T_{\sigma}:= T^\mu\!_{\sigma\mu}
\ee
is the trace of the torsion tensor.
It is immediate to see that this satisfies $S^\mu\!_{\nu\sigma}= -S^\mu\!_{\sigma\nu}$.

Let us now derive the Killing equation for a Riemann--Cartan spacetime.
The Killing equation gives a partial differential equation for vector fields generating isometries
\begin{equation}
\lie_{\xi} g_{\mu\nu} = 0\,.
\end{equation}

Explicitly,
\begin{equation}
\lie_{\xi} g_{\mu\nu} = \xi^{\rho}\partial_{\rho} g_{\mu\nu} + g_{\rho\nu}\partial_{\mu}\xi^{\rho} + 
g_{\rho\mu}\partial_{\nu}\xi^{\rho}\,.
\end{equation}

%
%

If we convert partial derivatives into covariant derivatives, namely
\ba
&&\partial_{\mu}\xi^{\nu} = \af_{\mu} \xi^{\nu} - \Gamma^{\nu}\!_{\mu \rho}\xi^{\rho}\,,\n\\
&&\partial_{\rho} g_{\mu\nu} = \af_{\rho} g_{\mu\nu} + 
\Gamma_{\nu\rho\mu} 
 +
 \Gamma_{\mu\rho\nu}\,,
\ea
it is straightforward to see that a Killing vector field $\xi$ satisfies
\begin{equation}\la{kill}
\lie_{\xi} g_{\mu\nu}=\af_{\mu} \xi_{\nu} + \af_{\nu} \xi_{\mu}  -
\xi^{\rho}\left( T_{\mu\nu\rho} +T_{\nu \mu\rho}     \right)
=0\,.
\end{equation}
In the case of the Levi--Civita connection, $T=0$,  we recover the standard $\lev_{(\mu}\xi_{\nu)}=0$ Killing equation.


Autoparallel curves on a Riemann--Cartan spacetime are defined through the equation
\begin{equation}\la{geodesic}
\xi^{\mu}\af_{\mu} \xi^{\nu} = -\tilde \kappa \xi^{\nu},
\end{equation}
in a generic parametrization where $\tilde \kappa$ is a measure of non-affinity. One should further note that, in Riemann--Cartan spacetime, $\tilde \kappa$ is a priori different from the surface gravity defined from the condition of the  Killing horizon generator to be a null vector field at the horizon (we elaborate more on this in Section \ref{sec:Hor}).

By means of the connection coefficient expression \eqref{ecconn}, the previous relation implies
\begin{equation}
\xi^{\mu}\lev_{\mu} \xi^{\nu} -T_{\mu\rho}\!^{\nu}\xi^{\mu}\xi^{\rho} = -\tilde \kappa \xi^{\nu}\,,
\end{equation}
where $\nabla$ is the covariant derivative w.r.t. the Levi--Civita connection.
Hence, autoparallel curves in a Riemann--Cartan spacetime are not extremal curves, since the latter is a notion defined with respect to the metric of the manifold and it yields  the standard geodesic equation in terms of the Levi--Civita connection only.

\section{non--Riemannian Local inertial frame}\la{sec:Inertial}


 The gravitational field strength is related to the components of a linear connection (which may be with or without torsion,  e.g.   the  Levi--Civita connection in Riemann  spacetime or the  Cartan connection  in Riemann--Cartan spacetime) in a local inertial frame.  Mathematically this means the existence of a unique local frame in which the connection components vanish at a point about which the local frame is described. The thermodynamical derivation strongly relies on the existence of such local inertial frame at each point of spacetime, in order to define a LCH in terms of a local Rindler horizon. 
 In this section we give a viable notion of the Einstein Equivalence Principle (EP) for a non--Riemannian spacetime having torsion.
 
In General Relativity one deals with a $n$-dimensional Riemannian  manifold denoted by $M_n$.  At each point $p$ of $M_n$, the tangent vector space is denoted as $T_p (M_n )$.  One can then introduce a local vector basis $e_a$. Given a local coordinate system  $\{x^{\mu} \}$, the frame $e_a$  is expanded  in terms of the local coordinate basis $\partial_{\mu}=\partial/\partial x^{\mu}$
\ba
e_a=e_a{}^{\mu}\partial_{\mu}\,.
\ea

For a $n$-dimensional Riemannian manifold $M_n$, the EP states that for any point $p$ a local inertial frame is introduced via a local coordinate transformation
\ba
dx^a \to dx^{\mu}(x^a(p))=e_a{}^{\mu}(x^b(p))dx^a
\ea
relating the flat Minkowski metric $\eta_{ab}$ to the induced metric,
 $g_{\mu\nu}=e^a{}_{\mu}e^b{}_{\nu} \eta_{ab}$,
of the curved Riemannian spacetime.

The coordinate basis is a set of $n$ linearly independent vectors, defined at each point of the manifold, which are tangent to the $n$ coordinate lines, which pass through that point and belong to a coordinate system (also called the global coordinate system or natural coordinate system) imposed on the manifold. For two different coordinate systems, the transformation between two coordinate bases can be defined as
\ba \la{holonomic}
e_{\mu}=e_{\mu}{}^{\nu}e_{\nu}\,,\quad {\rm where}\quad
 e_{a}{}^{\nu}=\partial x^{\mu}/\partial x^{a}\,;
 \ea
such transformations are integrable and called holonomic. They satisfy the condition 
\ba
\partial_{\nu} e_{a}{}^{\nu}-\partial_{\mu} e_{a}{}^{\mu}=0\,,
\ea
which corresponds to some integrability conditions for the coordinate system given by 
\ba
(\partial_{\mu} \partial_{\nu}-\partial_{\nu}\partial_{\mu})x^{a}=0\,.
\ea

In the Riemannian case, the $n^2 (n-1)/2$ integrability conditions effectively reduce the number of unknowns
$\partial_{\sigma}e_{j}{}^a$ from $n^3$  to $n^2 (n + 1)/2$.  This solves the system of $n^2 (n + 1)/2$ transformation equations
\ba
\Gamma^{\mu}_{\nu \rho}=e^{\mu}{}_a(e_{\nu}{}^b e_{\rho}{}^c \Gamma^a_{bc}-\partial_{\rho} e_{\nu}{}^a)=0\,,
\ea
thus providing the coordinate basis in which all the components of the Levi--Civita connection are locally set to zero at a point \cite{Gogala1980}.

We want to reproduce the argument for  Einstein--Cartan spacetime, that is to introduce a local frame in which the components of a more general affine connection (including torsion) vanish at some point $p$.  {The problem is, the connection $\Gamma(g, T)$ defined in \eqref{ecconn} can be set to zero,  under holonomic coordinate transformation \eqref{holonomic}, only if the torsion  tensor  vanishes  identically. Now more generally,  one can introduce a local inertial (Lorentz) frame in Einstein--Cartan spacetime by means of local vector basis $h_a$ defined by the non-integrable or anholonomic  \cite{schouten1954linear} transformations}
\ba
h_a=h_a{}^{\mu}(p)\partial_{\mu}\,,
\ea
with
\ba
\partial_{\nu}h^{a}{}_{\mu}-\partial_{\mu}h^{a}{}_{\nu} \neq 0\,.
\ea
This basis is also called a non-coordinate basis. Although it is always possible to find a coordinate basis which will coincide with an anholonomic basis locally, in one point, there is no coordinate system which would correspond to the anholonomic basis globally.

The tensor of anholonomicity is defined by
\ba
\Omega_{\nu \rho}{}^{\mu}=h_a{}^{\mu}(\partial_{\nu}h^{a}{}_{\mu}-\partial_{\rho}h^{a}{}_{\nu})\,.
\ea
With this basis, at every point, we define a local Lorentz frame by means of the set of coordinate differentials
\ba
dx^a=h^a{}_{\mu}(x^{\mu})dx^{\mu}\,.
\ea
Local Lorentz frames are then obtained by requiring the induced metric in these coordinates to be Minkowskian,
\ba
\eta_{ab}=h_a{}^{\mu}h_b{}^{\mu} g_{\mu \nu}\,.
\ea
Referred  to such an anholonomic system, the affine connection goes to
\ba
\Gamma^{\mu}_{\nu \rho} \to \Gamma^c_{ab}=h_a{}^{\mu}h_b{}^{\nu}h_{\rho}{}^{c}(T^{\mu}{}_{\nu \rho}-T_{\nu}{}^{\mu}{}_{\rho}+T_{\nu \rho}{}^{\mu}-\Omega^{\mu}{}_{\nu \rho}+\Omega_{\nu}{}^{\mu}{}_{\rho}-\Omega_{\nu \rho}{}^{\mu})\,.
\ea
At any arbitrary point $p$ of a 4-D spacetime, the orthonormality condition $g_{\mu \nu}=h_{\mu}{}^ah_{\nu}{}^b \eta_{ab}$ only determines 40 components of $\partial h$. The remaining 24 components may be locally fixed by $T(p) = \Omega(p)$. Then $\Gamma^c_{ab}(p)=0$ and, accordingly, torsion does not violate the EP. The local coordinate system $\{x^{\mu}\}$ associated with the anholonomic frame at $p$ can be extended to an infinitesimal neighborhood of the point $p$, in a similar way as a local normal coordinate basis is extended in the Riemannian case.

\section{Horizon properties}\la{sec:Hor}

For our derivation we are interested in a section of null surface, $\HH$. We now look at what happens to the result that a null surface is described by the geodesic flow of its normal vector field.

A null horizon surface can be defined by the implicit equation
\begin{equation}
\phi(x) = 0\,,
\end{equation}
with the condition that the normal vector field
\begin{equation}
\chi^{\mu} = h g^{\mu\nu}\partial_{\nu}\phi\,,
\label{normal}
\end{equation}
where $h$ is any non-vanishing function,
is null
\begin{equation}
\chi^{\mu}\chi_{\mu} = 0 \qquad \text{on } \HH\,.
\end{equation}
This last condition implies that the gradient of this norm is orthogonal to the surface $\HH$, \ie it is still proportional to the normal vector; namely 
the null normal to the horizon satisfies
\be\la{kappa}
\af_{\nu} (\chi^{\mu}\chi_{\mu}) = 2\chi^{\mu}\af_{\nu} \chi_{\mu}=-2\kappa \chi_{\nu}, \qquad \text{on } \HH\,,
\ee
with $\kappa$ a function corresponding to the normal surface gravity, which a priori represents a different notion of surface gravity than the inaffinity surface gravity defined through the geodesic equation \eqref{geodesic}.

For our horizon to be a Killing horizon, we further demand the null vector field $\chi^\mu$ to be a Killing vector field; we can then use \eqref{kill} to write
\ba
-\kappa \chi_{\nu}=\chi^{\mu}\af_{\nu} \chi_{\mu}=
-\chi^{\mu}\af_{\mu}\chi_{\nu} - 2 U^{\rho}\!_{\mu\nu}\chi^{\mu}\chi_{\rho}
=
-\chi^{\mu}\af_{\mu}\chi_{\nu}+T_{\mu\nu\rho}\chi^{\mu}\chi^{\rho}\,,
\ea
from which
\be
\chi^{\mu}\af_{\mu}\chi_{\nu}=\kappa \chi_{\nu}+T_{\mu\nu\rho}\chi^{\mu}\chi^{\rho}\,;
\label{eq:KillingFlow}
\ee
in terms of the Levi--Civita connection, the previous equation implies
\be
\chi^{\mu}\nabla_{\mu}\chi_{\nu}=\kappa \chi_{\nu}\,.
\ee
Hence, we see that the Killing vector generating the horizon is not geodesic with respect to the affine connection but rather with respect to the Levi--Civita connection associated to the metric (i.e the $\chi$ flows along extremal curves).

The non--geodesic flow of the Killing vector is quite a departure from the standard behavior of Killing horizons in General Relativity and indeed it does have striking consequences.   Let us first note that from \eqref{eq:KillingFlow} it is clear that the non-geodesic behavior is due to the presence of the term $T_{\mu\nu\rho}\chi^{\rho}\chi^{\mu}$, i.e.~of a torsion current across the horizon. In theories with non-propagating torsion this could be only carried on by a flow of particles which has an associated spin current. What is the effect of such a current on the horizon? (apart from inducing a non geodesic flow of the Killing vector)


An important observation in this sense is that for a non--vanishing tensor current across the horizon the above introduced definitions of surface gravity, the inaffinity $\tilde \kappa$ (defined by the geodesic equation \eqref{geodesic}) and the ``normal surface gravity" defined via \eqref{kappa}, do not coincide. This is also definitely at odd with what  one has in General Relativity where this (and others) definitions of the surface gravity do coincide for a stationary Killing horizon~\cite{Cropp:2013zxi}.

In order to show this explicitly let us conveniently write the ``normal surface gravity" as
\be
\kappa=-n^\nu\chi^\mu\af_\nu \chi_\mu\,,
\ee
where $n^\mu$ is an auxiliary null vector defined at the horizon such that $\chi^\mu n_\mu=-1$. 
Now, by means of the Killing equation \eqref{kill}, one has
\ba
\tilde \kappa&=& n^\nu\chi^\mu\af_\mu\chi_\nu\n\\
&=& - n^\nu\chi^\mu\af_\nu\chi_\mu+2n^\nu T_{(\mu\nu)\rho} \chi^\mu\chi^\rho\n\\
&=&\kappa+2n^\nu T_{(\mu\nu)\rho} \chi^\mu\chi^\rho\n
\,.
\ea

In GR the non coincidence of the inaffinity and normal definitions of the surface gravity for Killing horizon is generally associated to departure from equilibrium/stationarity, like for example in the case of evaporating/shrinking black holes (see e.g. the discussion in Section 2.2 of~\cite{Cropp:2013zxi} keeping in mind that the normal surface gravity basically coincides with the surface gravity notion associated to the near horizon peeling structure of outgoing light rays). In analogy, one might says that non-vanishing tensor currents across the horizon should not be allowed for a truly stationary description of the horizon and henceforth we shall ask them to be zero
\be\la{bingo}
T_{\mu\nu\rho}\chi^{\rho}\chi^{\mu}=0\, .
\ee

Remarkably, the above restriction on the torsion current across the horizon is crucial not only to remove ambiguities among otherwise inequivalent definitions of surface gravity, but also it is necessary in order to carry on the demonstration of the zeroth law of the black hole mechanics, i.e.~the proof of the constancy of the temperature across the horizon. In fact, in order to do so we first need to introduce a local set of null tetrads at the horizon, playing the role of coordinate vector fields. This construction can be implemented by means of Gaussian null coordinates. Since we need to use the Killing vector field as one of the coordinate vector field, we then need to impose \eqref{bingo} for it to be geodesic along the horizon (see Appendix \ref{App:C}).

Finally, we want to find the actual generator of the horizon, which is normally affinely parametrized on it. 
Let us then define on the horizon the vector field
\be\la{kchi}
k^\mu=e^{-\kappa v}\chi^\mu=\frac{1}{\kappa \lambda}\chi^\mu\,,
\ee
where $v$ is the non-affine parameter (Killing time) defined by
\be
\chi^\mu\af_\mu v=1\,.
\ee
We then have
\ba\la{geo}
k^{\mu}\af_{\mu}k_{\nu}&=&e^{-2\kappa v}(-\chi_\nu\chi^\mu\af_\mu(\kappa v) +\kappa \chi^\nu+T^\mu_{\nu\rho}\chi^{\rho}\chi_{\mu})\n\\
&=&
T^\mu\!_{\nu\rho}k^{\rho}k_{\mu}=\frac{1}{\kappa^2\lambda^2} T_{\mu \nu \rho}\chi^{\mu}\chi^{\rho}
\ea
As expected $k^\mu$ is an affinely parametrised null horizon generator only once the geodesic condition
\eqref{bingo} is imposed.

\section{Raychaudhuri equation}\la{sec:Ray}

The next ingredient in order to attempt a thermodynamical derivation of the Einstein--Cartan equation is the Raychaudhuri equation for a non--Riemannian spacetime. 
Let us thus proceed in its derivation.

We consider a local horizon $\HH$ generated by the affinely parametrised null vector $k^\mu$. 
Following the discussion of the above Section we shall demand the condition \eqref{bingo} to hold, hence \eqref{geo} becomes
\be
k^{\mu}\af_{\mu}k_{\nu}=0\,.
\ee

If we use this horizon null generator as an element of the horizon local coordinate basis and 
an auxiliary null vector field $n^\mu$ such that $k^\mu n_\mu=-1$, the spacetime metric can be decomposed as
\be\la{metric}
h_{\mu\nu}=g_{\mu \nu}+k_\mu n_\nu + k_\nu n_\mu\,,
\ee
where $h_{\mu\nu}$ is the transverse metric, namely
\be\la{kn}
k^\mu h_{\mu\nu}=n^\mu h_{\mu\nu}=0.
\ee

Let us denote $\eta^\mu$ as the deviation vector between two neighboring flux lines of $k^\mu$. The Lie derivative of $\eta^\mu$ along the tangent (to the horizon) vector   $k^\mu$ has to  vanish, namely
\be
\lie_{k} \eta^\mu=0\,.
\ee
This implies
\be
[k, \eta]^\mu=k^\nu\af_\nu \eta^\mu-\eta^\nu\af_\nu k^\mu-k^\sigma\eta^\nu T^\mu_{\sigma\nu} =0\,.
\ee
Thus, the failure of the deviation vector to be parallely transported along the horizon is measured by
\be
k^\nu\af_\nu \eta^\mu=\eta^\nu(\af_\nu k^\mu+k^\sigma T^\mu_{\sigma\nu})=\eta^\nu B^\mu\!_{\nu}\,,
\ee
where we have defined the deviation tensor
\be\la{B}
B_{\mu\nu}:= \af_\nu k_\mu+k^\sigma T_{\mu\sigma\nu}\,.
\ee
Let us now note that again the condition \eqref{bingo} is crucial as only in this case the deviation tensor \eqref{B} is, as usual, orthogonal to the generator vector field $k^\mu$, namely
\be\la{kB}
k^\mu B_{\mu\nu}=0= k^\nu B_{\mu\nu}\,,
\ee
where the first equality holds due to the null condition $k^{\nu}\af_{\mu}k_{\nu}=0$ while the second one can be deduced by the last line of \eqref{geo} and  \eqref{bingo}.

However, the deviation tensor above is not fully transversal since it is not orthogonal to $n^\mu$ and has a component along $n^\mu$. To obtain a purely transverse deviation tensor we can define the projection of $B_{\mu \nu}$ as
\ba\la{BB}
\tilde B_{\mu\nu}&=&h_\mu\!^\alpha h_\nu\!^\beta B_{\alpha \beta}\n\\
&=&B_{\mu\nu}+k_\mu n^\alpha B_{\alpha\nu}+k_\nu n^\alpha B_{\mu\alpha}+k_\mu k_\nu n^\alpha n^\beta B_{\alpha\beta}\,.
\ea

We can define the expansion of the congruence as 
\ba
\overline \theta &=&\frac{1}{2}h^{\mu\nu}\sL_k h_{\mu\nu}
=g^{\mu\nu}\tilde B_{\mu\nu}
=g^{\mu\nu} B_{\mu\nu}\,,
\ea
where we have used the orthogonality condition \eqref{kB} as well as \eqref{kn}.

Therefore, the expansion reads
\be\la{exp}
\overline \theta  =g^{\mu\nu}B_{\mu\nu}=\af_\nu k^\nu+k^\nu T_{\nu}\,.
\ee

The evolution of the expansion along $k^{\mu}$ (i.e along the horizon in our case) is given by
\ba
\frac{d\overline \theta }{d\lambda}&=&k^\mu\af_\mu\overline \theta = k^\mu\af_\mu \af_\nu k^\nu+  k^\mu\af_\mu(k^\nu T_{\nu})\n\\
&=& k^\mu\af_\mu \af_\nu k^\nu+  k^\mu k^\nu \af_\mu T_{\nu}+T^{\nu}T_{\mu\nu\rho}k^{\mu}k^{\rho}\,,
\ea
where the last term can be set to zero by using \eqref{bingo}, but we leave it  since we want to give the Raychaudhuri equation in full generality for a null congruence.

We can  expand the first term as
\ba
 k^\mu\af_\mu \af_\nu k^\nu&=&
k^\mu \af_\nu \af_\mu k^\nu-k^\mu \overline R_{\mu\nu \sigma}\!^\nu k^{\sigma}
-k^\mu T^\rho\!_{\mu\nu}\af_{\rho}k^{\nu}\n\\
&=&\af_\nu (k^\mu \af_\mu k^\nu)-\af_\nu k^\mu \af_\mu k^\nu
-\overline R_{\mu\sigma}k^\mu k^{\sigma}-k^\mu T^\rho\!_{\mu\nu}\af_{\rho}k^\nu\n\\
&=&\af^\nu (T^{\rho}\!_{\nu\mu} k^{\mu}  k_{\rho})
-\af_\nu k^\mu \af_\mu k^\nu
-\overline R_{\mu\sigma}k^\mu k^{\sigma}-k^\mu T^\rho\!_{\mu\nu}\af_{\rho}k^\nu\n\\
&=& (\af^\nu T^{\rho}\!_{\nu\mu} )k^{\mu}  k_{\rho}
+ T^{\rho}\!_{\nu\mu}\af^\nu (k^{\mu}  k_{\rho})
-\af_\nu k^\mu \af_\mu k^\nu
-\overline R_{\mu\sigma}k^\mu k^{\sigma}-k^\mu T^\rho\!_{\mu\nu}\af_{\rho}k^\nu\,.
\ea

Therefore
\ba\la{dtheta}
\frac{d\overline \theta }{d\lambda}
&=& -
k^\mu k^{\nu} \overline R_{\mu\nu}+ k^{\mu}  k_{\nu} \af^\sigma T^{\nu}_{\sigma\mu}+k^\nu k^\mu \af_\mu T_{\nu}\n\\
&-&\af_\nu k^\mu \af_\mu k^\nu+\underbrace{T^{\rho}_{\nu\mu}(\af^\nu k_{\rho}+\af_{\rho}k^\nu)k^{\mu} 
+T^{\rho}_{\nu\mu}(\af^\nu k^{\mu} ) k_{\rho}}_{=2\af_\mu k^\nu K^\mu\!_{\nu\sigma}k^\sigma}\n\\
&=&-k^\mu k^{\nu} \overline R_{\mu\nu}+ k^{\mu}  k_{\nu} \af^\sigma T^{\nu}_{\sigma\mu}+k^\nu k^\mu \af_\mu T_{\nu}
+T^{\sigma}T_{\mu\sigma\nu}k^{\mu}k^{\nu}\n\\
&-& B^{\nu\mu} B_{\mu\nu} + k^\rho T_{\nu\rho \mu}B_{\mu\nu}
+\af^\nu k^{\mu}(T_{\mu\nu\rho}+T_{\rho\nu\mu})k^{\rho}\,,
\ea
where in the last passage we have used the definition \eqref{B}.

One can show using \eqref{BB} and \eqref{bingo} that
\be
B^{\nu\mu} B_{\mu\nu}=\tilde B^{\nu\mu} \tilde B_{\mu\nu}\,.
\ee

Putting everything together, the Raychaudhuri equation reads
\ba\la{Ray}
\frac{d\overline \theta }{d\lambda}&=&
-\overline R_{\mu\nu}k^\mu k^{\nu}+ k^{\mu}  k_{\nu} \af^\sigma T^{\nu}_{\sigma\mu}+k^\nu k^\mu \af_\mu T_{\nu}
+T^{\sigma}T_{\mu\sigma\nu}k^{\mu}k^{\nu}\n\\
&-& \tilde B^{\nu\mu} \tilde B_{\mu\nu}
-\tilde B^{\mu\nu}(T_{\mu \rho\nu}-T_{\nu \rho\mu}+T_{\rho\mu\nu})k^\rho
+k^\sigma k^\rho T^\mu\!_{ \sigma}\!^\nu(T_{\mu \rho\nu}+T_{\rho\mu\nu})\,.
\ea

Written in terms of expansion, shear and twist, respectively
\ba
 \overline \theta &:=&h^{\mu\nu}\tilde B_{\mu\nu}\,,\\
 \overline\sigma_{\mu\nu}&:=&\tilde B_{(\mu\nu)}-1/2 \overline \theta  h_{\mu\nu}\,,\\
\overline\omega_{\mu\nu}&:=&\tilde B_{[\mu \nu]}\,,
\ea
the Raychaudhuri equation takes the form
\ba\la{Rayf}
\frac{d\overline \theta }{d\lambda}&=&
-\overline R_{\mu\nu}k^\mu k^{\nu}
- \frac{1}{2}\overline \theta^2-\overline\sigma^{\mu\nu}\overline\sigma_{\mu\nu}
+\overline\omega^{\mu\nu}\overline\omega_{\mu\nu}\n\\
&-&\overline\omega^{\mu\nu}(T_{\mu \rho\nu}-T_{\nu \rho\mu}+T_{\rho\mu\nu})k^\rho
+k^\sigma k^\rho T^\mu\!_{ \sigma}\!^\nu(T_{\mu \rho\nu}+T_{\rho\mu\nu})\n\\
&+& k^{\mu}  k_{\nu} \af^\sigma T^{\nu}\!_{\sigma\mu}+k^\nu k^\mu \af_\mu T_{\nu}
+T^{\sigma}T_{\mu\sigma\nu}k^{\mu}k^{\nu}\,.
\ea
Written in terms of the Riemannian expansion, shear and twist, the Raychaudhuri equation above can also be written as
\ba\la{Rayf2}
\frac{d\overline \theta }{d\lambda}&=&
-\overline R_{\mu\nu}k^\mu k^{\nu}
- \frac{1}{2} \theta^2-\sigma^{\mu\nu}\sigma_{\mu\nu}
+\omega^{\mu\nu}\omega_{\mu\nu}
+K^{\nu\mu}\!_\rho K_{\mu\nu\sigma} k^\rho k^\sigma\n\\
&+& k^{\mu}  k_{\nu} \af^\sigma T^{\nu}\!_{\sigma\mu}+k^\nu k^\mu \af_\mu T_{\nu}
+T^{\sigma}T_{\mu\sigma\nu}k^{\mu}k^{\nu}\,.
\ea

\section{entropy}\la{sec:Ent}

As for the Riemannian case, we can still assume  the entropy of a local causal horizon in Riemann--Cartan spacetime to be proportional to area. In fact, for a causal horizon at equilibrium the origin of its entropy is believed to be due to the vacuum entanglement across the horizon \cite{Bombelli:1986rw} (see, e.g. \cite{Pranzetti:2013lma, Oriti:2015rwa}, for a derivation of horizon entropy from the entanglement between quantum gravitational degrees of freedom) and, hence, the eventual presence of torsion due to some matter distribution away from the horizon is not expected to affect the proportionality of the entanglement entropy to the horizon area (at least not in a theory with non-propagating torsion, like Einstein--Cartan). 

Therefore, let us still assume an entropy-area law of the form
\be
S=\alpha A=\alpha \int dA=\alpha \int d^2x  \sqrt{h}\,,
\ee
where $A$ is area of horizon cross section, $h$ is determinant of the induced metric on the horizon cross section  and $\alpha $ is a proportionality constant which is generally dependent on the UV cut-off;
 the variation of this entropy (due to some physical process changing the horizon area) is given by
\be \la{s}
d S=\alpha \delta A= \int d^2x  \delta\sqrt{h} = \int d^2x \mathcal{L}_{k} \sqrt{h}\,.
\ee

By means of the relation
\be
\frac{d\sqrt{h}}{d \lambda}=\frac{1}{2}\sqrt{h}h^{\mu\nu}\frac{d h_{\mu\nu}}{d \lambda}\,,
\ee
 we obtain
 \ba
  \mathcal{L}_{k} \sqrt{h}=\xi^a\partial_a \sqrt{h}=\frac{1}{2}\sqrt{h}h^{\mu\nu}\frac{d h_{\mu\nu}}{d \lambda}=\frac{1}{2}\sqrt{h}h^{\mu\nu} \mathcal{L}_{k} h_{\mu\nu}\,,
 \ea
 where the last step is only true if we are working in Gaussian null coordinate.
 
 Using the metric decomposition \eqref{metric} in terms of the two null vectors and the projected metric, we have
 \ba \la{lieh}
 h^{\mu\nu} \mathcal{L}_{k} h_{\mu\nu}=h^{\mu\nu} \mathcal{L}_{k} [g_{\mu\nu}+k_\mu n_\nu+k_\nu n _a]=h^{\mu\nu} \mathcal{L}_{k} g_{\mu\nu}\,,
 \ea
 as $h^{\mu\nu} \mathcal{L}_{k}(k_\mu n_\nu)=0 $ due to the orthogonality of $k_\mu, n_\mu$ with $h^{\mu\nu}$.
Furthermore, one can show starting from \eqref{lieh} that
 \ba \la{lieg}
 h^{\mu\nu} \mathcal{L}_{k} g_{\mu\nu}&=&2h^{\mu\nu} (\af _\mu k_\nu)+T^\rho{}_{\sigma\mu}g_{\rho\nu}k^\sigma h^{\mu\nu}+T^\rho{}_{\sigma\nu}g_{\rho\mu}k^\sigma h^{\mu\nu}
 \n\\
 &=&2(\af^\mu k_\mu-T_{\mu \sigma\nu} k^\sigma n^\nu k^\mu+T_{\nu \sigma\mu }k^\sigma h^{\mu\nu})
 \n\\
 &=&2(\af^\mu k_\mu+T_{\nu \sigma\mu}k^\sigma g^{\mu\nu})\n\\
 &=&2(\af^\mu k_\mu+T_{ \mu}k^\mu )\,.
 \ea
 
 Finally using \eqref{lieg} in \eqref{s} we get the variation of entropy as,
\be
d S=  \int d\lambda d^2x  \sqrt{h}(\af^\mu k_\mu+T_{ \mu}k^\mu )=\int d\lambda d^2x  \sqrt{h}\, \overline \theta  \,,
\ee  
 where $\overline \theta$ is  the  expansion of the congruence in Riemann--Cartan spacetime, as defined in \eqref{exp}.

\section{Einstein--Cartan field equations as an equation of state}\la{sec:TEC}

We finally come to the derivation of the main result of this paper. Our objective is to start from thermodynamical variables, which ideally one can obtain as a result of coarse graining of spacetime, and show how the Einstein--Cartan field equation emerges after using the Clausius equation to relate them. For doing so we will use the variation of entropy we derived in previous section and the Raychaudhuri equation \eqref{Ray}. 

Before proceeding with the thermodynamical derivation, let us first recall the form of the Einstein--Cartan equation that we want to recover.

\subsection{Einstein--Cartan--Sciama--Kibble field equations}\la{sec:EC}
The Einstein--Cartan--Sciama-Kibble field equations read
\ba
&&\overline G_{\mu\nu}=8\pi G \left(T^{\va M}_{\mu\nu}+(\af_{\sigma}+T_{\sigma})(\tau^\sigma\!_{\mu\nu}-\tau_{\mu\nu}\!^\sigma-\tau_{\nu\mu}\!^\sigma)\right)\,,\la{eom1}\\
&& S^\sigma\!_{\mu\nu}=16 \pi G\, \tau ^\sigma\!_{\mu\nu}\la{eom2}\,,
\ea
where $T^{\va M}$ is the  metric (hence symmetric) stress-energy tensor (SET) containing also non--Riemannian contributions  and $\tau$ the spin angular momentum tensor. Given a matter Lagrangian $\sL_{\va M}$ depending only on the matter field $\psi$, its first derivatives $\af\psi$ and the metric $g$, these two quantities are defined as
\ba
T^{\va M}_{\mu\nu}:=\frac{2}{\sqrt{-g}}\frac{\delta \sL_{\va M}}{\delta g^{\mu\nu}}\,,\la{TM}\\
\tau^{\sigma}\!_{\mu}\!^{\nu}:=\frac{1}{2\sqrt{-g}}\frac{\delta \sL_{\va M}}{\delta K^{\mu}\!_{\sigma\nu}}\la{tau}\,.
\ea

The total action function yielding the Einstein--Cartan field equations above reads
\be
W=\int d^4x \left(\sL_{\va M}+\frac{1}{16\pi G} \sL_{\va G}\right)\,,
\ee
where the gravity Lagrangian reads
\be
\sL_{\va G}=\sqrt{-g} g^{\mu\nu}\overline  R_{\mu\nu}\,.
\ee

By combining the two Einstein--Cartan equations \eqref{eom1},  \eqref{eom2} we get 
\be\la{eom}
G_{\mu\nu}=8\pi G\, T^{\va M}_{\mu\nu}+\frac{1}{2}(\af_{\sigma}+T_{\sigma})(S^\sigma\!_{\mu\nu}-S_{\mu\nu}\!^\sigma-S_{\nu\mu}\!^\sigma)\,.
\ee
This is the equation we want to recover via the thermodynamical approach.

Let us start by splitting the equation \eqref{eom} into its symmetric and anti-symmetric parts and expand. The symmetric part of the Einstein--Cartan equation reads
\ba\la{eom-sym}
\overline R_{(\mu\nu)}-\frac{1}{2}g_{\mu\nu}(\overline R-2\Lambda)&=&8\pi G\, T^{\va M}_{\mu\nu}
-\frac{1}{2}(\af_{\sigma}+T_{\sigma})(S_{(\mu\nu)}\!^\sigma+S_{(\nu\mu)}\!^\sigma)\n\\
&=&8\pi G\, T^{\va M}_{\mu\nu}-\af^{\sigma} T_{(\mu\nu)\sigma}+\af_{(\mu} T_{\nu)} 
-T^\sigma T_{(\mu\nu)\sigma} +T_{(\mu} T_{\nu)}\,.
\ea

For later convenience, let us contract the field equation \eqref{eom} with two null vectors $k$; we find
\be\la{EC}
\overline R_{\mu\nu}k^\mu k^\nu=8\pi G\, T^{\va M}_{\mu\nu}k^\mu k^\nu+
\af^{\sigma} T^\mu\!_{\sigma\nu}k_\mu k^\nu+\af_\mu T_\nu k^\mu k^\nu
 +T^\sigma T_{\mu\sigma\nu}k^\mu k^\nu+T_\mu T_\nu k^\mu k^\nu
 \,.
\ee

The antisymmetric part of \eqref{eom} reads
\be\la{ECanti}
\overline R_{[\mu\nu]}=\frac{1}{2}(\af_{\sigma}+T_{\sigma})S^\sigma\!_{\mu\nu}\,.
\ee
However, this part of the Einstein--Cartan equations does not have a dynamical origin, but it follows simply from the definition of the Riemann tensor in terms of the connection \eqref{ecconn}; in fact, \eqref{ECanti} is equivalent to the Ricci tensor property \eqref{Ranti}, once the modified torsion tensor definition \eqref{S} is applied. Therefore, it is enough to recover the symmetric part of the Einstein--Cartan equations through the thermodynamical argument in order to capture their dynamical content. 

\subsection{Einstein--Cartan equation of state: Torsion as a geometric field }\la{sec:TEC1}

We are now ready to undertake the task we set for us at the start of this paper. Strong of the results of Section \ref{sec:Inertial}, we can construct at any point of a Riemann--Cartan spacetime a local inertial frame. In such a frame we can construct as in Section \ref{sec:RevE} a Rindler wedge within which local boost invariance  still has an associated  Unruh temperature, $T=\hbar/2\pi$, via the Bisognano--Wichmann theorem \cite{Haag:1992hx}. As we reviewed above, in presence of dissipative terms,  the Clausius law we shall enforce on such local Rindler wedge has to be generalized to take into account internal entropy terms; namely, one has to use the entropy balance law
\be\la{Cla}
dS=\frac{\delta Q}{T}+dS_i\,.
\ee
By means of the entropy formula \eqref{s}, we then have 
\be
 \alpha\delta A=\frac{2\pi}{\hbar}\delta Q+dS_i\,.
\ee

The heat flux across the LCH is given by the expression
\be\la{Q}
\delta Q=\int_\HH T^{\va M}_{\mu\nu}\chi^\mu d\Sigma^\nu
=- \int_\HH \sqrt{h} d\lambda d^2x \,\lambda\, T^{\va M}_{\mu\nu}k^\mu k^\nu\,,
\ee
where $d\Sigma^\nu=\sqrt{h} d\lambda d^2xk^\mu$ is the horizon volume element.

From the result of the previous Section, we have
\ba\la{dA}
\alpha \delta A
&=&\alpha \int_\HH \sqrt{h} d\lambda d^2x\, \overline \theta \n\\
&\approx& \alpha \int_\HH \sqrt{h}  d\lambda d^2x \left(\overline \theta _p +\lambda \left.\frac{d\overline \theta }{d\lambda}\right|_p\right)\,.
\ea
Therefore, using \eqref{dA} and \eqref{Q} in \eqref{Cla} the Clausius equation reads

\ba\la{dA2}
\alpha \int_\HH \sqrt{h} d\lambda d^2x \left(\overline \theta _p +\lambda \left.\frac{d\overline \theta }{d\lambda}\right|_p\right)=-\frac{2\pi}{\hbar} \int_\HH \sqrt{h} d\lambda d^2x\, \lambda\, T^{\va M}_{\mu\nu}k^\mu k^\nu+dS_i\,.
\ea
In the above equation, the l.h.s. has a  first term of zeroth order in $\lambda$ and a second term of first order in $\lambda$ , while  the r.h.s. is entirely first order in $\lambda$. Therefore in order to match the two sides of \eqref{Cla}, at the zeroth order in $\lambda $, we need the  condition $\theta_p=0$, which by \eqref{exp} implies
\be \la{eqcondition}
\af_\mu k^\mu=-T_\mu k^\mu|_p\,.
\ee

Now, before plugging in the Raychaudhuri equation \eqref{Rayf} in \eqref{dA2}, let us note that in presence of torsion, hypersurface orthogonality of the horizon generators does not imply anymore the vanishing of the null congruence twist. This is explicitly shown in Appendix \ref{App:B}. Therefore, in the general case, we can set neither the shear nor twist to zero in the Raychaudhuri eq. \eqref{Rayf} and they will contribute to the internal entropy term. 

In order to correctly identify all the contributions to this non-equilibrium term, we need to open up the non--Riemannian shear and twist in \eqref{Rayf}, which in general will  contribute terms both in $k^\mu$ as well as in its covariant affine derivatives, and identify only the latter as non-equilibrium contributions.
Hence, we use the explicit form of the horizon shear, twist and \eqref{eqcondition}, so to rewrite the Raychaudhuri equation \eqref{Rayf} as
\ba\la{Rayi}
 \left.\frac{d\overline \theta }{d\lambda}\right|_p&=&
-\overline R_{\mu\nu}k^\mu k^{\nu}+ k^{\mu}  k_{\nu} \af^\sigma T^{\nu}\!_{\sigma\mu}+k^\nu k^\mu \af_\mu T_{\nu}
+T^{\sigma}T_{\mu\sigma\nu}k^{\mu}k^{\nu}\n\\
&-&  \left. \af_\nu k^\mu \af_\mu k^\nu+2\af_\mu k^\nu K^\mu\!_{\nu\sigma}k^\sigma \right|_p
\,,
\ea
where the terms in the second line on the r.h.s. of \eqref{Rayi} are the internal entropy terms. 

Therefore, combining \eqref{dA2} with \eqref{Rayi}, the generalized Clausius law \eqref{Cla} implies at first order in $\lambda$
\be\la{Clai1}
- \frac{2\pi} {\hbar}T^{\va M}_{\mu\nu} k^\mu k^\nu\\
=\alpha
\left(
-\overline R_{\mu\nu}+\af^\sigma T^{\nu}\!_{\sigma\mu}+ \af_\mu T_{\nu}
+T^{\sigma}T_{\mu\sigma\nu} +T_\mu T_\nu \right)k^\mu k^{\nu}\,
\ee
and 
\ba\la{Clai2}
dS_i&=& \alpha   \int_\HH \sqrt{h}   d\lambda d^2x\,\lambda\left. \left(-\af_\nu k^\mu \af_\mu k^\nu+2\af_\mu k^\nu K^\mu\!_{\nu\sigma}k^\sigma -\af_\mu k^\mu \af_\nu k^\nu \right)\right|_p\n\\
&=&  \alpha   \int_\HH \sqrt{h}   d\lambda d^2x\,\lambda\left. \left(   -\sigma^{\mu\nu}\sigma_{\mu\nu}
+\omega^{\mu\nu}\omega_{\mu\nu}
+K^{\nu\mu}\!_\rho K_{\mu\nu\sigma} k^\rho k^\sigma -T_\mu T_\nu k^\mu k^{\nu}\right)\right|_p\,,
\ea
where, in the last equation, we have first included all the dissipative, non-equilibrium terms inside the Raychaudhuri equation (all the ones containing a covariant derivative of the horizon generator), and then re-expressed them, by means of \eqref{Rayf2}, \eqref{eqcondition}, in terms of the Riemannian shear and twist plus torsion contributions\footnote{Notice that the condition \eqref{eqcondition} induces an ambiguity in the identification of the equilibrium and the non-equilibrium parts of the Raychaudhuri, since the last term on the r.h.s. of  \eqref{Clai1} can always be compensated by a non-equilibrium one like the last one on the r.h.s. of the first line in \eqref{Clai2}. We have thus included them in order to consider the most general case.}.

The first relation \eqref{Clai1} yields, for
\be
\alpha=\frac{1}{4\hbar G}\,,
\ee
the symmetric part \eqref{EC} of the Einstein--Cartan equation.

The second one, eq. \eqref{Clai2}, provides a definition of the internal entropy contribution in presence of torsion. Notice that, for 
vanishing torsion (and hence twist), eq. \eqref{Clai2} reproduces the dissipative term obtained in \cite{Eling:2006aw, Chirco:2009dc}, namely
\be
dS_i= -\alpha  \int_\HH \sqrt{h}   d\lambda d^2x \,\lambda\,||\sigma||^2_p\,,
\ee
where the shear $\sigma_{\mu\nu}$ is the one defined w.r.t. the Levi--Civita connection. In the presence of torsion, the internal entropy  \eqref{Clai2} contains contributions coming from both the shear squared term and the twist terms inside the Raychaudhuri equation \eqref{Ray}. This could be considered as the generalisation to the Riemann--Cartan geometries of the Hartle--Hawking term describing the dissipation of a distortion of the horizon and it seems to imply a different output of gravitational waves w.r.t. to what expected in General Relativity (of course just in cases where at the horizons there are fluxes of matter generating torsion in non-propagating torsion theories like Einstein--Cartan).

As in the original argument of \cite{Jacobson:1995ab}, we could now try to use the Bianchi identity for a Riemann--Cartan spacetime in order to recover the Ricci scalar part of the equation of motion. However, in presence of torsion, the modified Bianchi identity contains torsion dependent terms which are not total covariant derivatives. Therefore, in this case, the modified Bianchi identity is of little help. 

However, one can split the symmetric part of the Einstein tensor  into a Riemannian part and a non--Riemanninan part (see next Subsection where this approach is carried out explicitly); the Riemannian term will be the standard Einstein tensor written in terms of the Levi--Civita connection and it will satisfy the standard Riemannian  spacetime Bianchi identity. The non--Riemanninan part of the symmetric Einstein tensor comprises terms involving the affine covariant derivative of the torsion and quadratic contractions of the torsion tensor. These terms can be moved to the r.h.s. of \eqref{eom-sym} in order to define an effective SET. 

The important point is that, on an Einstein--Cartan spacetime such SET will be conserved w.r.t.~the Levi--Civita connection, since the l.h.s.~is conserved due to the  Riemannian Bianchi identity. It follows that, on an Einstein--Cartan spacetime, the torsion tensor has to be such that the Levi--Civita covariant derivative of the non--Riemanninan part of the symmetric Einstein tensor has to be equal to the Levi--Civita covariant derivative of the r.h.s.~of \eqref{eom-sym}, namely

\ba\la{Bianchi2}
\nabla^\nu \overline R_{(\mu\nu)}-\frac{1}{2}\nabla_\mu(\overline R-2\Lambda)
=\nabla^\nu(8\pi G\, T^{\va M}_{\mu\nu}-\af^{\sigma} T_{(\mu\nu)\sigma}+\af_{(\mu} T_{\nu)} 
-T^\sigma T_{(\mu\nu)\sigma} +T_{(\mu} T_{\nu)})\,.
\ea
Once we split $\overline R_{(\mu\nu)}, \overline R$ into their Riemannian and non--Riemannian parts and use the Riemannian Bianchi identity\footnote{
The explicit splitting is obtained from
\ba
\overline R_{\mu\nu}&=&R_{\mu\nu}+\af_\sigma K^\sigma\!_{\mu\nu}-\af_\mu K^\sigma\!_{\sigma\nu} + K^\sigma\!_{\sigma\rho}K^\rho\!_{\mu\nu}-K^\sigma\!_{\mu\rho}K^\rho\!_{\sigma\nu}\n\\
&=&R_{\mu\nu}+\af_\sigma K^\sigma\!_{\mu\nu}+\af_\mu T_\nu -T_\rho K^\rho\!_{\mu\nu}+K^\sigma\!_{\rho\mu}K^\rho\!_{\sigma\nu}\,,\\
\overline R&=&g^{\mu\nu}\overline R_{\mu\nu}=R + 2g^{\mu\nu}\af_\sigma K^\sigma\!_{\mu\nu}+ g^{\mu\nu}(K^\sigma\!_{\sigma\rho}K^\rho\!_{\mu\nu}-K^\sigma\!_{\mu\rho}K^\rho\!_{\sigma\nu})\n\\
&=&R + 2\af^\mu T_\mu -T_\mu T^\mu - g^{\mu\nu}K^\sigma\!_{\mu\rho}K^\rho\!_{\sigma\nu}\,,\la{R}
\ea
where we have used the relations
\ba
K^\sigma\!_{\sigma\mu}&=&-T_\mu\,,\\
g^{\mu\nu}\af_\sigma K^\sigma\!_{\mu\nu}&=&-g^{\mu\nu}\af_\mu K^\sigma\!_{\sigma\nu}=\af^\mu T_\mu\,,\\
g^{\mu\nu}K^\sigma\!_{\sigma\rho}K^\rho\!_{\mu\nu}&=&-T_\mu T^\mu\,.
\ea},
 this represents a condition between torsion and the SET that characterizes an Einstein--Cartan spacetime, and thus it needs to be implemented in order to recover the Einstein--Cartan equation.

The Ricci scalar and cosmological constant parts in the symmetric Einstein--Cartan equation \eqref{eom-sym} can then be obtained from the condition \eqref{Bianchi2}, in  analogy to the standard Riemannian case. In fact, let us now go back to the part of the symmetric Einstein--Cartan equation that we recovered so far through the Clausius law, namely eq. \eqref{Clai1}. By peeling off the two $k$'s, we can rewrite this as
\be\la{EC2}
\overline R_{(\mu\nu)}
  +g_{\mu\nu}F(x)
=
8\pi G\,T^{\va M}_{\mu\nu}
-\af^{\sigma} T_{(\mu\nu)\sigma}+\af_{(\mu} T_{\nu)} 
-T^\sigma T_{(\mu\nu)\sigma} +T_{(\mu} T_{\nu)}\,,
\ee
where, as in the original thermodynamical derivation of  \cite{Jacobson:1995ab}, we have added a term proportional to the metric and depending on some function $F(x)$. By taking the Levi--Civita covariant derivative and enforcing the condition \eqref{Bianchi2}, it is then immediate to  obtain
\be
F(x)=-\frac{1}{2}\overline R +\Lambda\,.
\ee
Plugging this last relation back into \eqref{EC2} we thus recover the full symmetric Einstein--Cartan equation \eqref{eom-sym}.

\subsection{Einstein--Cartan equation of state: Torsion as a background field }

We now want to derive Einstein--Cartan equation from the non-equilibrium thermodynamical approach where we take the point of view of  torsion as an external (or background) field, with the torsion terms defining an effective SET  for a Riemannian spacetime. In fact,
if we write the symmetric part of the non--Riemannian Einstein tensor in terms of the Riemannian one plus torsion terms, namely
\ba\la{G}
\overline R_{(\mu\nu)}-\frac{1}{2}g_{\mu\nu}(\overline R-2\Lambda)&=&
 R_{\mu\nu}-\frac{1}{2}g_{\mu\nu}( R-2\Lambda) 
 -\af^\sigma T_{(\mu\nu)\sigma}+\af_{(\mu} T_{\nu)} +T^\sigma T_{(\mu\nu)\sigma}+K^\sigma\!_{\rho(\mu}K^\rho\!_{\sigma|\nu)}\n\\
& -&\frac{1}{2}g_{\mu\nu}\left( 2\af^\sigma T_\sigma -T_\sigma T^\sigma - g^{\alpha\beta}K^\sigma\!_{\alpha\rho}K^\rho\!_{\sigma\beta}\right)\,,
\ea
then the symmetric part of the Einstein--Cartan equation \eqref{eom-sym} can be written in terms of the Riemannian Einstein tensor and an effective SET, namely
\ba\la{ECR}
 R_{\mu\nu}-\frac{1}{2}g_{\mu\nu}( R-2\Lambda)&=&
 8\pi G\, T^{\va M}_{\mu\nu} 
-2T^\sigma T_{(\mu\nu)\sigma} +T_{(\mu} T_{\nu)}-K^\sigma\!_{\rho(\mu}K^\rho\!_{\sigma|\nu)}\n\\
&+&\frac{1}{2}g_{\mu\nu}\left( 2\af^\sigma T_\sigma -T_\sigma T^\sigma - g^{\alpha\beta}K^\sigma\!_{\alpha\rho}K^\rho\!_{\sigma\beta}\right)\,.
\ea

As the next step, we rewrite the Raychaudhuri equation \eqref{Rayf2} in terms of the Riemannian Ricci tensor; this yields
\ba\la{Rayf3}
\frac{d \theta }{d\lambda}&=&
- R_{\mu\nu}k^\mu k^{\nu}
- \frac{1}{2} \theta^2-\sigma^{\mu\nu}\sigma_{\mu\nu}
+\omega^{\mu\nu}\omega_{\mu\nu}
-2T^{\sigma}T_{(\mu\nu)\sigma}k^{\mu}k^{\nu}\,.
\ea
We can now run the non-equilibrium thermodynamical argument, similarly to the previous Subsection. By means of the generalized Clausius relation \eqref{Cla}, at first order in $\lambda$, it is immediate to see that the EC equation written as in \eqref{ECR}, modulo the terms proportional to the metric $g_{\mu\nu}$, is recovered once we use exactly the same definition of internal entropy production term as in the previous derivation (namely, the second line of \eqref{Clai2}); explicitly, we recover 
\be\la{Clai1a}
- \frac{2\pi} {\hbar}T^{\va M}_{\mu\nu} k^\mu k^\nu\\
=\alpha
\left(
- R_{\mu\nu}
-2T^{\sigma}T_{(\mu\nu)\sigma}  
+T_{(\mu} T_{\nu)}-K^\sigma\!_{\rho(\mu}K^\rho\!_{\sigma|\nu)}\right)k^\mu k^{\nu}\,
\ee
for
\ba\la{Clai2a}
dS_i
&=&  \alpha   \int_\HH \sqrt{h}   d\lambda d^2x\,\lambda\left. \left(   -\sigma^{\mu\nu}\sigma_{\mu\nu}
+\omega^{\mu\nu}\omega_{\mu\nu}
+K^{\nu\mu}\!_\rho K_{\mu\nu\sigma} k^\rho k^\sigma -T_\mu T_\nu k^\mu k^{\nu}\right)\right|_p\,.
\ea
The Ricci scalar part of the equation of motion can be recovered 
similarly like in the previous derivation, by means of the Riemannian Bianchi identity. In the present case, this approach is even more well justified  
given that we have explicitly expressed the EC equation in terms of the  Riemannian Einstein tensor. 

This second approach to the thermodynamical derivation of the EC equation, with all the non--Riemannian torsion contributions reabsorbed in an effective SET, may seem more linear and clean at first. However, if we had proceeded along these lines from the beginning, the definition of the internal entropy production term \eqref{Clai2a} would have appeared as an ad hoc one, in order to recover the desired result. On the other hand, in the derivation presented in Subsection \ref{sec:TEC1}, where we work with the geometrical structures of the Riemann--Cartan spacetime, this form of $dS_i$ follows naturally from the Raychaudhuri equation \eqref{Rayf2}.

\section{discussion}\la{sec:Disc}

In this paper we have extended the thermodynamics of spacetime formalism to the Einstein--Cartan theory of gravity. 
In doing so we have had to reconsider several ingredients entering in the original derivation~\cite{Jacobson:1995ab}. First by redefining the notion of the local inertial frame in Riemann--Cartan geometries as well as by reconsidering the notion of Killing horizon surface gravity and rederiving the Raychaudhuri equation in this framework. In doing so we have obtained several original results and we have understood that the notion of a stationary Killing horizon in this setting requires an additional condition on the torsion current through the horizon which enforces the geodesic flow of the Killing vector and the horizon generator.

Then we have applied this toolkit to the spacetime thermodynamics approach and shown that, using a generalised Clausius equation, it is possible to recover the relevant part of the Einstein--Cartan--Sciama--Kibble equations. In doing so we have identified the relevant non-equilibrium terms, finding a novel dependence on the twist, which thus represent a generalisation of the usual Hartle--Hawking term. Let us stress that this term is calculated on the horizon and as such can present non-zero twist and torsion even in theories with non-propagating torsion as Einstein--Cartan, as long as the deformation of the local Rindler horizon is generated by matter fluxes endowed with spin currents.
This seems to suggest that if these terms are as usual associated to the actual energy that can be observed at infinity as carried away by gravitational waves, then they can provide a signature of the actual presence of torsion in the case e.g.~of black hole mergers in environments with suitable matter.


\section{acknowledgment}

The authors wish to thank Goffredo Chirco, 	
Christopher Eling, Daniele Oriti, Lorenzo Sindoni, Sebastiano Sonego and Vincenzo Vitagliano for discussions at the early stage of this project, and  Goffredo Chirco and 
Christopher Eling  for their feedback on a final version of this manuscript as well as.
We acknowledge the John Templeton Foundation for the supporting grant \#51876. 


\appendix
\section{Properties of Riemann--Cartan spacetime}\la{App:A}

We review here some other  properties of Riemann--Cartan spacetime important for the derivation of the results in the main part of the text.

The commutator of the covariant derivatives in presence of torsion is given by
\ba\la{comm}
[\af_{\mu},\af_{\beta}]k^{\nu}&=& \partial_{\mu}(\af_{\beta}k^{\nu})-\Gamma^{\rho}_{\mu\beta}\af_{\rho}k^{\nu}+ 
\Gamma^{\nu}_{\mu\rho}\af_{\beta}k^{\rho} -\mu\leftrightarrow\beta\n\\
&=&(\partial_{\mu}\Gamma^{\nu}_{\beta\sigma}-\partial_{\beta}\Gamma^{\nu}_{\mu\rho}+\Gamma^{\nu}_{\mu\rho}\Gamma^{\rho}_{\beta\sigma}-\Gamma^{\nu}_{\beta\rho}\Gamma^{\rho}_{\mu\sigma})k^{\sigma}
-(\Gamma^{\rho}_{\mu\beta}-\Gamma^{\rho}_{\beta\mu})\af_{\rho}k^{\nu}\n\\
&=&-\overline R_{\mu\beta\sigma}\!^\nu k^{\sigma}-T^\rho_{\mu\beta}\af_{\rho}k^{\nu}\,,
\ea
where we have used the standard definition of the Riemann tensor, although expressed in terms of the Riemann--Cartan connection \eqref{ecconn}.
 Notice that from this definition of the Riemann tensor the following symmetry properties still hold:
\ba
&&\overline R_{\mu\rho\sigma}\!^\nu=-\overline R_{\rho\mu\sigma}\!^\nu\,,\la{R1}\\
&&\overline R_{\mu\rho\sigma}\!^\alpha g_{\alpha \nu}=\overline R_{\mu\rho\sigma\nu}=-\overline R_{\mu\rho\nu\sigma}=-\overline R_{\mu\rho\nu}\!^\alpha g_{\alpha \sigma}.\la{R2}
\ea
The first relation (\ref{R1}) is obvious from the definition (\ref{comm}) of $\overline R_{\mu\beta\sigma}\!^\nu$; the second (\ref{R2}) can be shown by applying the commutator (\ref{comm}) to the metric $g_{\mu\nu}$ and using the metricity condition to eliminate the extra terms proportional to the torsion. However, the other symmetry property does not hold anymore, namely
\be
\overline R_{\mu\rho\sigma\nu}\neq \overline R_{\sigma\nu\mu\rho};
\ee
this is the case since the property $\overline R_{[\mu\rho\sigma]}\!^\nu=0$ is no longer true in the presence of torsion. As a consequence, the Ricci tensor is not symmetric anymore. More precisely, it can be shown that
\be\la{Ranti}
\overline R_{\mu\nu}=\overline R_{\nu\mu}-3\af_{[\mu} T^\sigma\!_{\sigma\nu]}+T_\sigma T^\sigma\!_{\mu\nu}\,.
\ee

\section{Hypersurface orthogonal congruence in presence of torsion }\la{App:B}

In this Appendix we show that, in presence of torsion, hypersurface orthogonality does not imply a vanishing twist. 
If we consider a surface defined  by the implicit equation $\phi(x)=0$, the vector field normal to the surface is given by 
\ba
\chi_{\mu}=h\partial_{\mu}\phi\,,
\ea
where $h$ is a proportionality constant.   In Riemann--Cartan geometry, the modified commutator \eqref{comm}  implies, for a generic function $f$,
\ba
\af_{[\mu}\af_{\nu]}f=-T^{\rho}{}_{\mu\nu}\af_{\rho}f
\ea
and thus the Frobenius theorem takes the form 
\ba \label{frobenius}
\chi_{[\mu}\af_{\nu}\chi_{\rho]}=-\chi_{\mu}T^{\sigma}{}_{\nu \rho}\chi_{\sigma}-\chi_{\nu}T^{\sigma}{}_{\rho\mu}\chi_{\sigma}-\chi_{\rho}T^{\sigma}{}_{\mu \nu}\chi_{\sigma}\,.
 \ea 
In the Riemannian case ($T=0$), the r.h.s. of \eqref{frobenius} vanishes and the resulting relation can be used to prove that the twist of the congruence vanishes as well. If we now try to reproduce the standard proof in presence of torsion, we see that this is no longer necessarily the case. More precisely, 
by means of \eqref{BB}, for the horizon generator null vector $k$, we have
\ba
k_{[\mu} \omega_{\nu \rho]}=k_{[\mu} \tilde B_{\nu \rho]}=k_{[\mu} B_{\nu \rho]}+k_{[\mu}B_{\nu |\sigma |}k_{\rho]}n^{\sigma}+k_{[\mu}B_{\sigma|\rho }k_{\nu]}n^{\sigma}\,;
\ea
it is straightforward to see that the last two terms vanish, so we can write 
\ba
k_{[\mu} B_{\nu \rho]}&=&k_{[\mu} \af_{\rho}k_{\nu]}+k_{[\mu}T_{\nu|\sigma|\rho]}k^{\sigma}\nonumber\\
&=&k_{[\mu} \af_{\rho}k_{\nu]}+k_{\mu}T_{\nu \sigma \rho}k^{\sigma}-k_{\nu}T_{\mu \sigma \rho}k^{\sigma}+k_{\nu}T_{\rho \sigma \mu}k^{\sigma}-k_{\rho}T_{\nu \sigma \mu}k^{\sigma}+k_{\rho}T_{\mu \sigma \nu}k^{\sigma}-k_{\mu}T_{\rho \sigma \nu}k^{\sigma}\,.
\ea
Therefore,  the condition for hypersurface orthogonality expressed as eq. \eqref{frobenius}, where we can clearly replace $\chi$ with $k$ given their relation \eqref{kchi}, does no longer imply $k_{[\mu} B_{\nu \rho]}=0$ and, hence, in general $\omega_{\mu\nu}\neq 0$.


\section{Zeroth law}\la{App:C}
In this Appendix we prove that the normal surface gravity defined by \eqref{kappa} provides a good notion of horizon temperature even in the case of non-vanishing torsion, namely it satisfies the zeroth law of horizon thermodynamics. 

In order to do so we introduce a set of null tetrads made by the Killing vector (using the geodesic condition \eqref{bingo}) plus a second null geodesic vector $n^\mu$ and a complex null vector $m^\mu$ tangent to the horizon 2-sphere cross-section such that
\be\la{norm}
\chi^\mu n_\mu=-1=-m^\mu \bar m_\mu~~~  {\rm and}~~~ n^\mu n_\mu=m^\mu m_\mu=\bar m^\mu \bar m_\mu=0.
\ee
In this coordinate system adapted to the the null surface, these four null vector represent  well-defined basis and the
 pull-back of the metric on the 2-sphere can be written as
\be
h_{\mu\nu}=m^{(\mu}\bar m^{\nu)}=g_{\mu \nu}+\chi_\mu n_\nu + \chi_\nu n_\mu
\ee
and
\be
\chi^\mu h_{\mu\nu}=n^\mu h_{\mu\nu}=0\,; 
\ee
moreover, being elements of a coordinate basis, the null vectors satisfy
\be\la{comms}
[k,n]^\mu=[k,m]^\mu=[n,m]^\mu=0\,.
\ee

We are now ready to compute the Lie derivative along the horizon Killing vector field generator of $\kappa$.
From \eqref{kappa}, we can write
\be\la{kappaex}
\kappa=-n^\rho\chi^\mu\af_\rho \chi_\mu\,.
\ee
We then have
\ba
\lie_\chi \kappa&=&\chi^\nu\af_\nu\kappa=- \chi^\nu\af_\nu(n^\rho\chi^\mu\af_\rho \chi_\mu)\n\\
&=&- \chi^\nu n^\rho\chi^\mu\af_\nu\af_\rho \chi_\mu
+\kappa\chi_\rho \chi^\nu \af_\nu n^\rho  
- \chi^\nu n^\rho\af_\nu\chi^\mu\af_\rho \chi_\mu\n\\
&=&- \chi^\nu n^\rho\chi^\mu\af_\rho  \af_\nu\chi_\mu
+\chi^\nu n^\rho\chi^\mu R_{\nu\rho\sigma\mu}\chi^\sigma
+\chi^\nu n^\rho\chi^\mu T^\sigma_{\nu\rho}\af_{\sigma} \chi_\mu\n\\
&+&\kappa\chi_\rho \chi^\nu \af_\nu n^\rho  
- \chi^\nu n^\rho\af_\nu\chi^\mu\af_\rho \chi_\mu\n\\
&=&- \chi^\nu n^\rho\chi^\mu\af_\rho  \af_\nu\chi_\mu
-\kappa \chi^\nu n^\rho\chi_\sigma T^\sigma_{\nu\rho}\n\\
&+&\kappa\chi_\rho \chi^\nu \af_\nu n^\rho  
- \chi^\nu n^\rho\af_\nu\chi^\mu\af_\rho \chi_\mu\,,
\ea
where in the last step we have used the symmetry properties of the Riemann tensor, the definition \eqref{kappa} and the commutator \eqref{comm}.
We now compute
\ba
- \chi^\nu n^\rho\chi^\mu\af_\rho  \af_\nu\chi_\mu&=&
-\af_\rho(\chi^\nu n^\rho\chi^\mu  \af_\nu\chi_\mu)
+(n^\rho\af_\rho\chi^\nu+\chi^\nu\af_\rho n^\rho)\chi^\mu\af_\nu \chi_\mu
+\chi^\nu n^\rho\af_\nu\chi^\mu\af_\rho\chi_\mu\n\\
&=&
-2\af_\rho(\chi^\nu n^\rho\chi^\mu  \af_{(\nu}\chi_{\mu)})
-\kappa n^\rho\chi_\nu\af_\rho\chi^\nu
+\chi^\nu n^\rho\af_\nu\chi^\mu\af_\rho\chi_\mu\n\\
&=&
-\af_\rho\left(n^\rho\chi^\nu \chi^\mu \chi^{\sigma} \left(T_{\mu\nu\sigma}+T_{\nu\mu\sigma}\right)\right)
-\kappa n^\rho\chi_\nu\af_\rho\chi^\nu
+\chi^\nu n^\rho\af_\nu\chi^\mu\af_\rho\chi_\mu\n\\
&=&
-\kappa n^\rho\chi_\nu\af_\rho\chi^\nu
+\chi^\nu n^\rho\af_\nu\chi^\mu\af_\rho\chi_\mu\,,
\ea
where in the last passage we have used the Killing eq. \eqref{kill}.

Therefore,
\ba
\lie_\chi \kappa&=&
-\kappa \chi^\nu n^\rho\chi_\sigma T^\sigma_{\nu\rho}
-\kappa n^\rho\chi_\nu\af_\rho\chi^\nu
+\kappa\chi_\rho \chi^\nu \af_\nu n^\rho\n\\
&=&\kappa \chi_\rho [\chi, n]^\rho\n\\
&=&0\,,
\ea
where we have used the definition
\be
[\chi, n]^\mu=\chi^\nu\af_\nu n^\mu-n^\nu\af_\nu \chi^\mu-\chi^\sigma n^\nu T^\mu\!_{\sigma\nu}
\ee
and the property \eqref{comms}.

To show the surface gravity is constant we need to further show $h_{\alpha}\!^{\nu}\af_{\nu}\kappa=0$.
If we assume the existence of bifurcation surface $S_0$ at which $\chi^\mu=0$ we can show $h_{\alpha}\!^{\nu}\af_{\nu}\kappa=0$ at the bifurcation surface as follows 
\ba
h_{\alpha}\!^{\nu}\af_{\nu}\kappa|_{\va S_0}&=&h_{\alpha}\!^{\nu}\af_\nu(n^\rho\chi^\mu\af_\rho \chi_\mu)|_{\va S_0}\n\\
&=&- h_{\alpha}\!^{\nu} n^\rho\chi^\mu\af_\nu\af_\rho \chi_\mu
- h_{\alpha}\!^{\nu} \chi^\mu\af_\nu n^\rho\af_\rho \chi_\mu
- h_{\alpha}\!^{\nu} n^\rho\af_\nu\chi_\mu\af_\rho \chi^\mu|_{\va S_0}\n\\
&=&- h_{\alpha}\!^{\nu} \af_\nu\chi_\mu n^\rho \af_\rho \chi^\mu|_{\va S_0}\n\\
&=&- h_{\alpha}\!^{\nu} \af_\nu\chi_\mu \left(\chi^\rho \af_\rho n^\mu
-\chi^\sigma n^\rho T^\mu\!_{\sigma\rho}\right)|_{\va S_0}\n\\
&=&0\,,
\ea
where we have used the fact that the horizon Killing vector field commutes with the auxiliary null vector $n^\mu$, as well as the vanishing of  $\chi^\mu$ at the bifurcation surface.

Now one can show that

\ba
\mathcal{L}_{\chi}(h_{\alpha}\!^{\nu}\af_{\nu}\kappa)=h_{\alpha}\!^{\nu}\mathcal{L}_{\chi}(\af_{\nu}\kappa)+\mathcal{L}_{\chi}(h_{\alpha}\!^{\nu})\af_{\nu}\kappa=0\,.
\ea
In order to do so, let us notice first that
the use of Gaussian null coordinates adapted to the horizon implies $\mathcal{L}_{\chi}(h_{\alpha}\!^{\nu})=0$ (as follows from \eqref{comms}); furthermore,
\ba
h_{\alpha}\!^{\nu}\mathcal{L}_{\chi}(\af_{\nu}\kappa)&=&h_{\alpha}\!^{\nu}(\chi^{\mu}\af_{\mu}\af_{\nu}\kappa+\af_{\mu}\kappa \af_{\nu}\chi^{\mu}+\chi^{\mu}\af_{\sigma}\kappa\, T^{\sigma}{}_{\mu \nu})
\n\\
&=&h_{\alpha}\!^{\nu}(\chi^{\mu}\af_{\nu}\af_{\mu}\kappa-\chi^{\mu} T^{\sigma}{}_{\mu \nu}\af_{\sigma}\kappa+\af_{\mu}\kappa \af_{\nu}\chi^{\mu}+\chi^{\mu}\af_{\sigma}\kappa T^{\sigma}{}_{\mu \nu})
\n\\
&=&h_{\alpha}\!^{\nu}(\af_{\nu}(\chi^{\mu}\af_{\mu}\kappa)-\af_{\nu}\chi^{\mu}\af_{\mu}\kappa+\af_{\mu}\kappa \af_{\nu}\chi^{\mu})
\n\\
&=&0\,,
\ea
where we used the constancy of $\kappa$ along the horizon Killing vector field, as we derived earlier.
 Therefore, $h_{\alpha}\!^{\nu}\af_{\nu}\kappa$ is constant over the horizon and thus if it is $0$ at one point, it will be $0$ everywhere.


\end{document}